\documentclass[lettersize,journal]{IEEEtran}
\IEEEoverridecommandlockouts
\usepackage{graphicx, amsmath, amsthm, amssymb, subcaption, url, cite, array, amsthm,booktabs,xcolor}
\usepackage[ruled, linesnumbered]{algorithm2e} 
\usepackage{algorithm2e, setspace}
\usepackage[font={small}]{caption}
\usepackage{hyperref}

\usepackage{bm}
\usepackage{threeparttable}
\usepackage{pifont}
\usepackage{amssymb} 
\usepackage{multirow}

\usepackage{xcolor}

\usepackage{lipsum}  
\IEEEoverridecommandlockouts
\makeatletter
\let\@oldmaketitle\@maketitle
\renewcommand{\@maketitle}{
  \@oldmaketitle
  \vspace{-15pt}  
}
\makeatother

\begin{document}

\title{\huge Multi-Modal Beamforming with Model Compression and Modality Generation for V2X Networks}

\author{Chen Shang,~\IEEEmembership{Graduate Student Member,~IEEE}, Dinh Thai Hoang,~\IEEEmembership{Senior Member,~IEEE},\\ Jiadong Yu,~\IEEEmembership{Member,~IEEE}
\thanks{
Chen Shang is with the Internet of Things Thrust, The Hong Kong University of Science and Technology (Guangzhou), Guangzhou, China. He is now with the Faculty of Engineering and Information Technology, University of Technology Sydney, Sydney, Australia (e-mail: chen.shang@student.uts.edu.au). Dinh Thai Hoang is with the School of Electrical and Data Engineering, University of Technology Sydney, Australia (hoang.dinh@uts.edu.au). Jiadong Yu is with the Internet of Things Thrust, The Hong Kong University
of Science and Technology (Guangzhou), Guangzhou, Guangdong, China (jiadongyu@hkust-gz.edu.cn) }}
\maketitle
\begin{abstract}
Integrating sensing and communication (ISAC) is a promising technology for predictive beamforming in 6G vehicle-to-everything (V2X) networks. However, current ISAC paradigms rely solely on radio-frequency (RF)-based sensing, which limits sensing resolution and beamforming robustness in complex wireless environments. Fortunately, the widespread deployment of diverse non-RF sensors such as cameras and LiDAR, along with the integration of artificial intelligence (AI) and communication systems, offers new opportunities to improve the synergy between sensing and communication.
Motivated by this, this work develops a multi-modal sensing-assisted beamforming framework for realistic V2X scenarios. Specifically, we propose BeamTransFuser, a hierarchical Transformer-based multi-modal learning framework that exploits cross-modal correlations among camera, LiDAR, radar, and GPS observations to improve beam prediction accuracy and robustness. To facilitate practical deployment on roadside units, we further develop a module-aware pruning scheme to reduce inference latency while preserving  competitive performance. Furthermore, to address potential missing-modality conditions in real-world scenarios, we introduce a generative model that is able to reconstruct missing inputs from available observations, allowing the framework to operate reliably even under incomplete sensing conditions.
Extensive experimental results conducted on real-world datasets demonstrate that the proposed scheme consistently outperforms existing baselines across various metrics.
\end{abstract}

\section{Introduction}

\subsection{Background}
Over the past decades, wireless communication systems have undergone continuous and transformative evolution. To support the next stage of development, IMT-2030 has identified six enabling technologies for future sixth-generation (6G) systems~\cite{ITU2023}. Among these, integrated sensing and communication (ISAC) holds significant promise. By integrating wireless communication and radar sensing into a unified framework, ISAC significantly reduces signaling overhead, shortens beam alignment latency, and improves spectral efficiency~\cite{11358925,10960485}. These advantages make ISAC particularly well-suited for emerging applications such as vehicle-to-everything (V2X) networks~\cite{5GAA_P190033,11091493} and autonomous driving~\cite{10944644}.

Despite its significant advantages, ISAC still faces considerable challenges in practical deployment. Specifically, existing sensing-centric, communication-centric, and joint design approaches~\cite{9737357} each involve inherent trade-offs. Sensing-centric and communication-centric designs typically prioritize one function over the other, which may compromise communication throughput, standard compatibility, sensing resolution, or adaptability. Joint design approaches aim to balance both functions, but typically incur higher system complexity and implementation difficulty, making practical deployment more difficult~\cite{9737357}.
Furthermore, a common limitation of these approaches is their heavy dependence on radio-frequency (RF)-based sensing. In complex vehicular environments, especially under dense infrastructure, blockage, high mobility, and non-line-of-sight (NLoS) conditions, RF signals are highly susceptible to multipath propagation and signal degradation, which can severely reduce sensing accuracy and beam alignment reliability. These effects pose significant challenges to practical ISAC deployment in real-world V2X scenarios~\cite{10944644,10955337}.

Fortunately, the widespread deployment of diverse sensors such as cameras and LiDAR opens new opportunities to address above challenges. As non-RF sources, these sensors hold strong potential to complement RF-based ISAC by providing rich contextual and structural information that RF sensors alone cannot capture.
For example, LiDAR provides precise three-dimensional (3D) geometric information and high range accuracy by capturing the spatial structure and depth of the surrounding environment, which is crucial for accurate localization and obstacle detection~\cite{9127813}. Cameras, on the other hand, provide dense semantic and texture information, enabling tasks such as object classification, lane detection, and traffic sign recognition~\cite{9127813}. In addition, unlike RF-based sensors, these non-RF sensors are less affected by multi-path fading and signal blockage, which often degrade RF signal quality in cluttered or NLoS environments. This makes them particularly effective in complex and rapidly changing urban scenarios, where rich visual semantics and structural awareness are crucial for reliable sensing. 

As a result, leveraging these diverse sensors to enhance ISAC helps overcome the limitations of traditional solely RF-based designs by providing more robust environmental awareness, thereby enabling stronger synergy between sensing and communication, i.e., multi-modal sensing for communication~\cite{10330577}.

\subsection{Related Works and Challenges}
\begin{table*}[t]
\centering
\footnotesize
\caption{Comparison of representative multi-modal beamforming methods. $\clubsuit$ indicates the functionality is not explored, \ding{55} denotes violation of the latency requirement (5GAA TR P-170142~\cite{5GAA_P190033}), and \checkmark indicates explicit support.}
\label{tab:related_work_comparison}
\begin{tabular}{>{\centering\arraybackslash}m{3.8cm} >{\centering\arraybackslash}m{3.8cm} >{\centering\arraybackslash}m{3.8cm} >{\centering\arraybackslash}m{3.8cm}}
\toprule
\textbf{Method} & \textbf{Sensing Modalities} & \textbf{RSU-Side Deployment Feasibility} & \textbf{Robustness to Missing Modalities} \\
\midrule
Position-based~\cite{10683225} & GPS & $\clubsuit$ & $\clubsuit$ \\
TII~\cite{tian2023multimodal} & Camera, GPS & $\clubsuit$ & $\clubsuit$ \\
CMDF~\cite{10.1145/3653644.3680497} & Camera, Radar & \ding{55} & $\clubsuit$ \\
ICMFE~\cite{10912462} & Camera, Radar & \ding{55} & $\clubsuit$ \\
QTNs~\cite{10577431} & Camera, LiDAR, GPS & $\clubsuit$ & $\clubsuit$ \\
MMDL~\cite{10949645} & Camera, LiDAR, GPS & \checkmark & $\clubsuit$ \\
Multi-modal~\cite{10683225} & Camera, LiDAR, Radar, GPS & $\clubsuit$ & $\clubsuit$ \\
MMT~\cite{10735366} & Camera, LiDAR, Radar, GPS & \ding{55} & $\clubsuit$ \\
\textbf{Ours (BeamTransFuser)} & \textbf{Camera, LiDAR, Radar, GPS} & \textbf{\checkmark} & \textbf{\checkmark} \\
\bottomrule
\end{tabular}
\end{table*}

Different from traditional ISAC works that mainly focus on physical-layer issues such as waveform/signal design~\cite{10549948,11044619,11169758} and sensing-communication resource allocation~\cite{9945983,11554053}, multi-modal sensing-assisted communication exploits heterogeneous sensing information as side information to support communication tasks such as beam prediction~\cite{10330577,DeepSense2}.
To fully realize the potential of these multi-modal sensors, it is essential to extract and fuse their modality-specific semantic features that are closely related to communication and sensing tasks. Compared with RF signals, which can be efficiently processed using traditional signal processing techniques, non-RF sensing data usually contain richer spatial and semantic information and therefore require advanced methods such as deep learning to capture complex representations. Therefore, designing efficient deep learning frameworks to process these multi-modal data is crucial for fully unlocking their potential.

Benefiting from another key enabling technology of 6G, i.e., the integration of artificial intelligence (AI) and communication~\cite{ITU2023}, several recent works have leveraged advanced AI technologies to achieve the aforementioned objectives. For instance,~\cite{tian2023multimodal} and~\cite{10577431} designed Transformer-based frameworks~\cite{vaswani2017attention} to deeply extract semantic information from different modalities, aiming to enhance the beamforming performance via these multi-modal sensing data. Alternatively,~\cite{10912462,10949645,10735366} proposed an advanced deep learning framework with cross-modal feature enhancement mechanisms to further improve the robustness and accuracy of beam prediction. 
However, most existing methods~\cite{tian2023multimodal,10577431} only consider a limited subset of modalities (e.g., camera and GPS), leaving other valuable modalities such as LiDAR and radar underutilized. Although a few schemes~\cite{10683225,10735366} attempt to incorporate a broader range of sensors (e.g., camera, LiDAR, radar, and GPS), their performance remains limited due to the absence of a unified and well-designed deep learning architecture that can effectively exploit the full potential of multi-modal sensing data.

Furthermore, the aforementioned methods primarily focus on improving beamforming prediction accuracy, while overlooking the significant computational and communication overhead introduced by large and complex AI models. To be specific, although these models achieve high performance by leveraging multiple architectural modules, their extensive resource demands and high inference latency impose considerable pressure on edge devices such as roadside units (RSUs) in V2X networks. While RSU-side deployment can be supported by more stable power supply and edge computing resources, the latency-sensitive nature of V2X services still requires high inference efficiency for the deployed beam prediction model under practical edge conditions. In addition, all of these methods~\cite{tian2023multimodal,10.1145/3653644.3680497,10577431,10912462,10949645,10735366} overlook the realistic scenario where some modalities may be unavailable, and thus exhibit limited robustness to missing modality inputs. In other words, once a model is trained with a fixed input configuration, the absence of any expected modality can lead to input dimension mismatches, rendering the model incapable in practical deployment~\cite{liang2025aligning}.

Given the above discussions and the systematic comparison in TABLE~\ref{tab:related_work_comparison}, we summarize the following three key challenges that remain insufficiently addressed in the current literature:
\begin{itemize}
\item First, given the heterogeneous nature of multi-modal sensing data (e.g., camera, LiDAR, radar, and GPS) and the rapidly changing vehicular environment, how can the system effectively fuse these diverse modalities to support accurate and adaptive beamforming decisions? This requires not only unified representation learning across spatially and semantically distinct modalities, but also fusion strategies that are robust to sensor noise and modality inconsistency.

\item Second, given the latency constraints of beamforming tasks and the computational constraints of RSUs, how can complex multi-modal deep learning models be efficiently deployed for real-time inference? This issue is further complicated by the fact that high-performing models typically employ large-scale architectures and multi-stage fusion pipelines, which introduce significant inference latency and render them unsuitable for delay-sensitive communication tasks.

\item Third, how can robust beamforming be ensured when sensor inputs are missing due to environmental disturbances, sensor failures, or the absence of certain sensing modalities? 
This challenge is particularly critical in real-world deployments where sensor availability cannot be guaranteed, requiring the system to adaptively compensate for missing inputs without compromising performance or requiring retraining.

\end{itemize}

\subsection{Motivation and Contributions}
To address the above challenges, this work proposes a robust and adaptive multi-modal beamforming system for real-world V2X networks. Specifically, we first design an end-to-end beamforming framework, termed \textbf{BeamTransFuser}, which employs modality-specific branches to extract features from each sensing source (i.e., camera, LiDAR, radar, and GPS) and incorporates a hierarchical multi-stage Transformer-based fusion backbone to progressively model cross-modal dependencies across different representation levels. This architecture enables more accurate and context-aware beamforming decisions in dynamic vehicular environments. 
In addition, beyond improving beam prediction accuracy, we further consider the latency constraints of real-time communication tasks and propose a splitting-based pruning strategy that compresses the model into a lightweight version suitable for edge deployment. This strategy customizes the pruning process for different modules based on their architectural characteristics, enabling substantial parameter reduction while maintaining performance. Such compression significantly reduces inference latency and computational load, making the system more suitable for real-time deployment. Finally, to ensure robustness under missing modality conditions, we incorporate a generative model that reconstructs absent modality features based on the available inputs. This enhances the model's adaptability by bridging the gap between partial observations and the expected input of BeamTransFuser, thereby improving system reliability in practical deployment scenarios with incomplete sensing conditions. To the best of our knowledge, this is the first work that employs multi-modal sensing data for beamforming while explicitly considering practical deployment constraints, including real-time inference, edge resource limitations, and incomplete sensor observations.
Our main contributions are summarized as follows:
\begin{itemize}
    \item 
    We propose an end-to-end deep learning framework BeamTransFuser, which leverages heterogeneous multi-modal sensing data to enable intelligent beamforming in vehicular networks.
    By jointly reasoning over all available modalities, the proposed architecture significantly improves beam prediction accuracy in dynamic vehicular environments.
    
    \item We develop a splitting-and-pruning-based model compression strategy that enables efficient deployment of BeamTransFuser on resource-limited RSUs. The compressed model significantly reduces computational overhead while maintaining accuracy, achieving near real-time inference across different hardware platforms.

    \item We introduce a generative model to address the potential challenge of missing modality. By leveraging the correlations among different modalities, the generative model is able to reconstruct missing modalities from the available ones, enabling BeamTransFuser to maintain robust beamforming performance even under incomplete sensing conditions without requiring re-training.

    \item We conduct extensive experiments using real-world multi-modal V2X datasets to evaluate the effectiveness of the proposed BeamTransFuser framework.
    The results demonstrate that our method consistently outperforms existing approaches~\cite{a2022_deepsense,10683225,tian2023multimodal,10.1145/3653644.3680497,10577431,10912462,10949645,10735366} in terms of beamforming performance, highlighting its strong generalization capability and practical applicability in dynamic vehicular scenarios.
\end{itemize}

The rest of our paper is organized as follows.
Section~\ref{system overview} introduces the system model and formulates the problem.
Section~\ref{BeamTransFuser Framework} presents the architecture of BeamTransFuser in detail.
In Section~\ref{se:enhancing}, we describe the proposed model compression strategy and the generative model.
Section~\ref{simultion} provides comprehensive experimental results to evaluate the performance of the proposed framework.
Finally, Section~\ref{conclusion} concludes the paper.

\section{System Overview And Problem Formulation}\label{system overview}
\begin{figure}[t]
    \centering
    \includegraphics[width=0.98\linewidth]{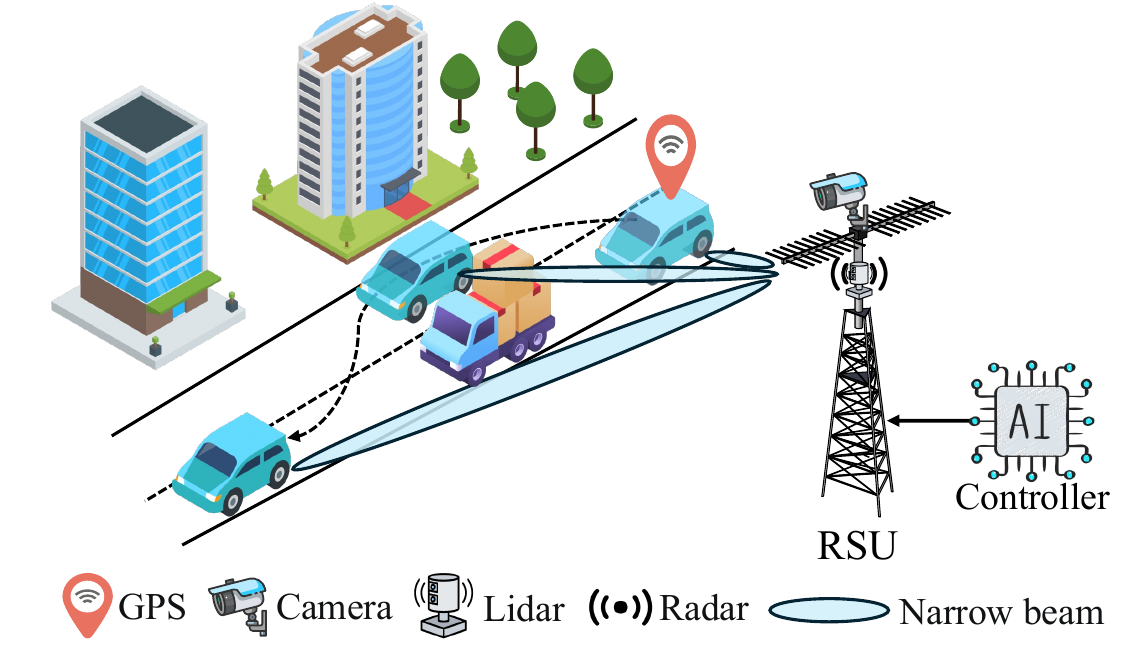}
    \caption{Multi-modal sensing-assisted communication system. The operation process of the proposed system includes
(1) The sensors deployed at the RSU collect real-time sensing data, (2) the collected data are fed into a pretrained deep learning model deployed at the RSU, and (3) based on the input data, the RSU generates the corresponding beamforming scheme.}
    \label{fig:scenario}
    \vspace{-18pt}
\end{figure}

As shown in Fig.~\ref{fig:scenario}, we consider a multi-modal sensing-assisted communication system, where an RSU is equipped with a uniform linear array for signal transmission and reception.
Unlike conventional ISAC systems that rely solely on a single sensing modality (i.e., radar), our system integrates multiple types of sensors like GPS, camera, LiDAR, and radar to capture real-time environmental information. These heterogeneous modalities can complement each other, enabling robust beamforming decisions under varying environmental and vehicular conditions. For example, when the communication channel between the RSU and the vehicle is in an NLoS, the radar sensor may fail to detect the target reliably due to signal blockage or reflection. Such sensing limitations can be compensated by supplementary information from other sensors.

We consider that the RSU generates the beamforming vector based on a predefined beamforming codebook, denoted by $\mathcal{F} =\left\{ \mathbf{f}_m \right\} _{m=1}^{M}$, where each $\mathbf{f}_m$ is a complex-valued column vector, and $M$ is the total number of beamforming candidates. Let $\mathbf{f}_m$ be the transmit beamforming vector used to transmit the downlink signal $s$. Accordingly, the received communication signal at the vehicle, denoted by $\mathcal{C}$, can be expressed as:
\begin{equation}
\mathcal{C} = \mathbf{h}^{\mathrm{H}} \mathbf{f}_m s + z_c
\label{eq:communication for vehicle},
\end{equation}
where $\mathbf{h}$ denotes the downlink channel vector from the RSU to the vehicle, and $\mathbf{h}^{\mathrm{H}}$ is its conjugate transpose (Hermitian transpose). In addition, $z_c \sim \mathcal{CN}(0, \sigma^2)$ represents circularly symmetric complex Gaussian noise.

As a result, the signal-to-noise ratio (SNR) of the communication link, denoted by $\gamma$, is given by:
\begin{equation}
\gamma(\mathbf{f}_m) = \frac{\left| \mathbf{h}^{\mathrm{H}} \mathbf{f}_m \right|^2}{\sigma^2}.
\label{eq:snr}
\end{equation}
Accordingly, the achievable transmission rate can be expressed as:
\begin{equation}
R= \log_2\left(1 + \gamma_n(\mathbf{f}_m)\right).
\label{eq:rate}
\end{equation}

In practice, the maximum transmission rate can be achieved by selecting the optimal beamforming vector $\mathbf{f}^*$ from all candidates in $\mathcal{F}$, which can be expressed as:
\begin{equation}
\mathbf{f}^* = \underset{\mathbf{f}_m \in \mathcal{F}}{\arg\max} \; \log_2 \left( 1 + \gamma(\mathbf{f}_m) \right).
\label{eq:optimal-beam}
\end{equation}
However, this work instead considers multi-modal sensing-assisted beam prediction, where the goal is to exploit heterogeneous sensing observations to infer the optimal beam index from real-world data. Accordingly, we develop a learning-based framework for beam index prediction. Notably, accurate beam index prediction directly leads to better beam alignment, thereby enhancing the received signal quality and ultimately improving the overall communication performance.

\begin{figure*}[t]
    \centering
    \includegraphics[width=0.98\linewidth]{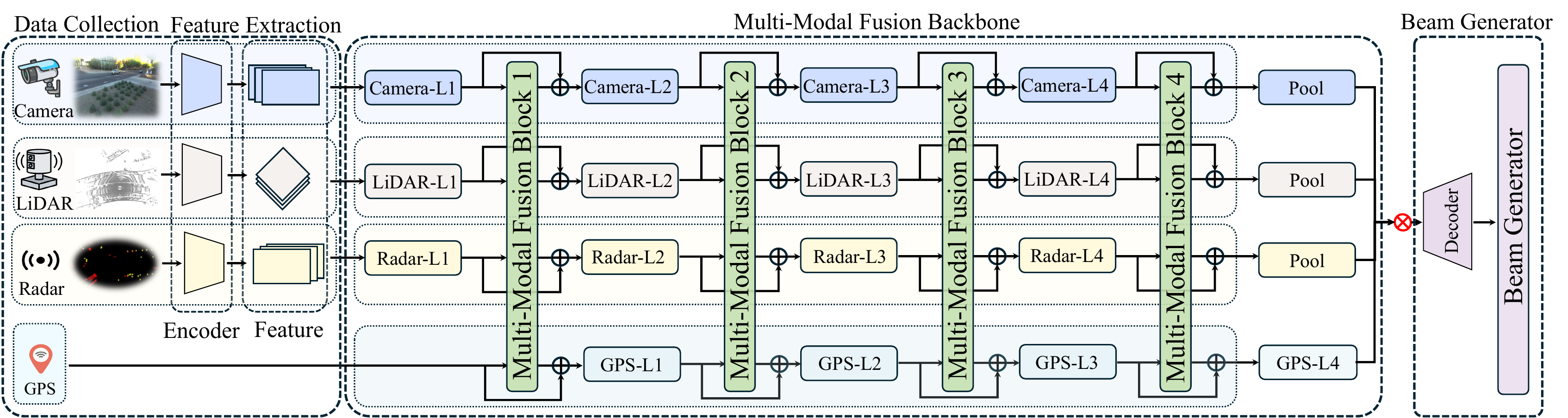}
    \caption{The overview of BeamTransFuser. BeamTransFuser consists of four modality-specific branches (i.e., camera, LiDAR, radar, and GPS), followed by a hierarchical Transformer-based Multi-Modal Fusion backbone composed of four stages of modality-specific layers (e.g., Camera-L1$\rightarrow $Camera-L2$\rightarrow $Camera-L3$\rightarrow $Camera-L4), with Multi-Modal Fusion Blocks interleaved between stages to progressively integrate cross-modal features. Residual connections ($\bm{\oplus}$) between backbone stages help preserve modality-specific information and facilitate gradient flow.
    After the final stage, the output features from each modality are globally pooled and fused via a learnable softmax-weighted aggregation mechanism (${\color{red} \bm{\otimes}}$). The fused representation is then processed by the Beam Generator to predict the optimal beam index.}
    \label{Fig.BeamTransFuser}
\end{figure*}

Let $\mathcal{M}_{\boldsymbol{\Theta}}(\cdot)$ denote the deep learning model, where $\boldsymbol{\Theta}$ denotes the model parameter vector (including all weights and biases), and $(\cdot)$ represents the input data, such as red-green-blue (RGB) images, radar signals, and LiDAR point cloud data.
Specifically, we consider a dataset $\mathbf{\mathcal{X}} = \{\mathcal{X}_i\}_{i=1}^N$ consisting of $N$ multi-modal sensing samples collected from real-world sequential data streams (e.g., camera, LiDAR, radar, and GPS), along with a corresponding label set $\mathbf{\mathcal{Y}} = \{\mathcal{Y}_i\}_{i=1}^N$, where each label $\mathcal{Y}_i$ denotes the index of the beamforming vector in $\mathcal{F}$ that achieves the maximum transmission rate for the $i$-th sample.
In other words, $\mathcal{Y}_i$ represents the optimal beam index for the $i$-th sample.

As a result, the objective is to learn a deep neural model $\mathcal{M}_{\boldsymbol{\Theta}}(\cdot)$ that maps each multi-modal input $\mathcal{X}_i$ to the corresponding beam index $\mathcal{Y}_i$, thereby approximating the optimal beam selection that maximizes the transmission rate in~\eqref{eq:rate}. Let $\boldsymbol{\Theta}^*$ denote the optimal model parameters, the learning task can be formulated as the following optimization problem:
\begin{equation}
\textbf{P1:}~ \boldsymbol{\Theta}^* = \underset{\boldsymbol{\Theta}}{\arg\min}~ \sum_{i=1}^{N} \mathcal{L} \left( \mathcal{M}_{\boldsymbol{\Theta}}(\mathcal{X}_i), \mathcal{Y}_i \right),
\label{eq:beamtransfuser}
\end{equation}
where $\mathcal{L}(\cdot)$ denotes the loss function used to measure the discrepancy between the predicted beam indices and the ground truth labels, e.g., cross-entropy loss~\cite{mao2023cross} and focal loss~\cite{lin2017focal}.

The main challenge in solving \textbf{P1} lies in designing a deep learning model $\mathcal{M}_{\boldsymbol{\Theta}}(\cdot)$ that can effectively process heterogeneous multi-modal inputs and accurately predict the optimal beam index. This requires the model to extract latent semantic features, capture inter-modal dependencies, and learn robust mappings from raw sensor data to beamforming decisions.
These demands pose significant challenges for conventional architectures such as convolutional neural networks (CNNs) and long short-term memory (LSTM) networks~\cite{DeepSense_Challenge}, which often lack the capacity to model complex cross-modal interactions. 
To address these challenges, in the following section, we present \textbf{BeamTransFuser}, a multi-modal fusion network that efficiently combines modality-specific branches with hierarchical Transformer modules to enable robust and adaptive beamforming through deep cross-modal understanding.

\section{BeamTransFuser Framework}\label{BeamTransFuser Framework}

As illustrated in Fig.~\ref{Fig.BeamTransFuser}, the proposed BeamTransFuser framework consists of four modality-specific branches corresponding to camera, LiDAR, radar, and GPS data, respectively. Each branch is composed of multiple specialized modules designed to extract features at different levels. While modality-specific branches extract features individually, effective cross-modal understanding requires integrating complementary information across modalities. To this end, four multi-modal fusion blocks are interleaved respectively throughout the network to progressively aggregate and align features (e.g., multi-modal fusion block 1 between Layer 1 and Layer 2). Each fusion block fuses complementary features across modalities, allowing the model to learn a more comprehensive and robust representation of the environment.
In addition, residual connections (i.e., $\oplus$) are inserted between adjacent layers to preserve modality-specific information and facilitate gradient flow during backpropagation. This design enhances the model's training stability and accelerates convergence.

The fused features are then fed into a learnable softmax-weighted aggregation module (i.e., ${\color{red} \otimes }$), which dynamically assigns weights to different modalities based on their relevance to the beamforming task. This allows the model to adaptively focus on the most informative modalities under different conditions. Subsequently, the fused feature representation passes through a decoder network followed by a beam generator, which predicts the optimal beam from a predefined codebook.

In the following, we first provide a detailed explanation of the data processing pipelines and network architectures used in the four modality-specific branches. Then, we introduce the design of the multi-modal fusion block, which models the cross-modal dependencies to support robust beamforming decisions.

\vspace{-8pt}
\subsection{Multi-Modal Branches}

We present the modality-specific branches for camera, LiDAR, radar, and GPS in this subsection. As shown in Fig.~\ref{Fig.BeamTransFuser}, except for GPS, each branch consists of a modality-specific encoder that transforms raw sensor data into initial spatial representations. These encoded features are then fed into the shared multi-modal fusion backbone, which incorporates feature extractors and hierarchical Transformer blocks to capture cross-modal dependencies. Note that although all branches share a consistent architectural pattern, their channel dimensions and configurations are customized to fit the characteristics of each sensor modality.

\subsubsection{Multi-Modal Encoders}
To transform raw data into spatially structured features, we adopt a modality-specific convolutional encoder for each input stream. Each encoder consists of a convolutional layer followed by batch normalization, ReLU activation, and max pooling. This unified architectural design serves as a shallow feature extractor that projects heterogeneous inputs into a shared spatial grid with 64 output channels, thereby enabling consistent processing in subsequent fusion stages.

Specifically, the camera encoder processes RGB images and produces 64-channel feature maps, while the LiDAR and radar encoders handle 1-channel and 2-channel projections, respectively. Furthermore, since GPS data is low-dimensional and structured as textual or tabular information (e.g., [\textit{latitude}, \textit{longitude}]), it does not require a dedicated encoder like those used for image or point cloud data. Instead, it is directly projected into the shared feature space via a multilayer perceptron (MLP) for downstream fusion.

\subsubsection{Multi-Modal Feature Extraction}

Following the encoder, each modality branch is equipped with a pretrained ResNet-based feature extractor comprising four hierarchical residual stages (i.e., Layer~1 to Layer~4). To accommodate the heterogeneity among sensor modalities, we adopt a modality-aware backbone configuration. In particular, ResNet34~\cite{he2016deep} is assigned to the camera branch to extract rich semantic and spatial representations from high-resolution RGB images, whereas a lightweight ResNet16~\cite{he2016deep} variant is employed for the LiDAR and radar branches, which typically contain lower-dimensional or sparse data. This design ensures that each modality is processed with an appropriate model capacity, mitigating overfitting risks for low-information inputs while reducing overall computational cost.

To facilitate progressive and multi-scale fusion, multi-modal fusion blocks (detailed in Section~\ref{Fusion Block}) are interleaved between residual stages. At each level, feature maps from all modalities are average-pooled into fixed-size grids, projected into a shared embedding space, and concatenated into a unified \textit{token} sequence. This \textit{token} sequence is then processed by a shared multi-head self-attention module, enabling fine-grained cross-modal interactions. The fused features are subsequently upsampled and reintegrated into each modality branch via residual connections, enabling bidirectional information exchange while preserving modality-specific representations. This progressive fusion strategy allows the network to model both low-level and high-level inter-modal dependencies while maintaining spatial structure across modalities throughout the pipeline.

\vspace{-10pt}
\subsection{Multi-Modal Fusion Block}\label{Fusion Block}
As illustrated in Section~\ref{system overview}, this work aims to leverage multi-modal sensing data to assist the beamforming process. However, utilizing data from diverse modalities introduces several significant challenges. 
On the one hand, the inherent heterogeneity among sensing sources, including differences in spatial resolution, data formats, and information density, makes direct feature alignment and fusion difficult. More importantly, each modality encodes distinct semantic representations of the environment~\cite{10330577}. For instance, RGB images provide rich textures and color information but lack depth perception, LiDAR captures accurate yet sparse 3D geometric structures, and radar yields coarse signals with limited semantic content. 
These semantic gaps hinder the extraction of shared contextual representations and effective cross-modal reasoning. Naive fusion strategies, such as direct concatenation, often fail to capture modality-specific characteristics, resulting in redundant or suboptimal representations. Therefore, a carefully designed fusion mechanism is essential to exploit complementary strengths while suppressing noisy or irrelevant information.

To address these challenges, we adopt the Transformer~\cite{vaswani2017attention} architecture to construct the multi-modal fusion block. By leveraging the multi-head self-attention mechanism (MHSA), Transformers can effectively model complex dependencies across different input modalities. This allows the model to dynamically attend to the most informative features and learn cross-modal interactions within a unified representation space~\cite{9716741}.
Note that since our task focuses on beamforming rather than complex reasoning tasks such as natural language processing, we only utilize the encoder-style self-attention architecture of the original Transformer for multi-modal fusion~\cite{vaswani2017attention}. Its details are illustrated on the left side of Fig.~\ref{fig:pruning}.

Specifically, the Transformer operates on a sequence of tokens, which serve as the basic units of input representation~\cite{vaswani2017attention}. In our task, these tokens are derived from modality-specific feature representations after backbone feature extraction and spatial normalization. At each fusion stage, the feature maps from the camera, LiDAR, and radar branches are first normalized to a common spatial resolution using adaptive average pooling with a fixed $8\times 8$ anchor grid. The pooled feature maps are then flattened into token sequences, so that all spatial modalities share the same token granularity before entering the Transformer encoder. For the GPS modality, since it does not possess a native 2D spatial structure, it is projected by a stage-specific linear embedding layer and incorporated into the unified token sequence as non-spatial tokens. In this way, the joint attention space is established through shared anchor-based spatial alignment for the spatial modalities and stage-wise embedding-dimension consistency across all modalities. Unlike modality-specific attention modules, our design concatenates all modality tokens into a unified sequence and applies a shared multi-head self-attention mechanism. After self-attention, the fused tokens are reorganized according to modality and passed back to the corresponding branches for subsequent processing. This design enables efficient and implicit cross-modal interaction, reduces parameter redundancy, and allows the model to dynamically attend to informative tokens across all modalities within a single attention space. We detail the multi-modal fusion block as follows.

Let $\mathbf{F} \in \mathbb{R}^{\mathcal{N} \times \mathcal{D}}$ denote the unified token sequence obtained by concatenating features from all sensor modalities, where $\mathcal{N}$ is the total number of \textit{tokens} and $\mathcal{D}$ is the feature embedding dimension. The shared \textit{query}, \textit{key}, and \textit{value} matrices, denoted by $\mathbf{Q}$, $\mathbf{K}$, and $\mathbf{V}$, respectively, are computed as:
\begin{equation}
\mathbf{Q} = \mathbf{F} \mathbf{W}^{\mathrm{Q}}, \quad
\mathbf{K} = \mathbf{F} \mathbf{W}^{\mathrm{K}}, \quad
\mathbf{V} = \mathbf{F} \mathbf{W}^{\mathrm{V}},
\label{eq:dimension for embedding}
\end{equation}
where $\mathbf{W}^{\mathrm{Q}}, \mathbf{W}^{\mathrm{K}},~\text{and}~ \mathbf{W}^{\mathrm{V}} \in \mathbb{R}^{\mathcal{D} \times \mathcal{D}}$ are learnable projection matrices shared across all modalities. These matrices project the unified token features into three distinct subspaces, enabling the Transformer to model token-level relationships across different modalities. By computing attention weights based on the similarity between queries and keys, the model selectively aggregates complementary information from the value vectors. This attention-based interaction refines each token representation, allowing it to incorporate relevant context from other modalities and thereby enhancing cross-modal feature integration~\cite{vaswani2017attention}. For notational simplicity, we omit the explicit head index in the following formulation.

Accordingly, the attention scores are computed via the scaled dot-product attention mechanism~\cite{vaswani2017attention}:
\begin{equation}
\mathrm{Scores}(\mathbf{Q}, \mathbf{K}) = \frac{\mathbf{Q} \mathbf{K}^{\top}}{\sqrt{\mathcal{D}/h}},
\end{equation}
where $h$ is the number of attention heads.
These scores are normalized using the softmax function to obtain the attention weights:
\vspace{-2pt}
\begin{equation}
\mathrm{Attention}(\mathbf{Q}, \mathbf{K}) = \mathrm{softmax} \left( \mathrm{Scores}(\mathbf{Q}, \mathbf{K}) \right).
\vspace{-3pt}
\end{equation}
The fused token representation, denoted by $\tilde{\mathbf{F}} \in \mathbb{R}^{\mathcal{N} \times \mathcal{D}}$, is then obtained by applying the attention weights to the value matrix~\cite{vaswani2017attention}:
\vspace{-2pt}
\begin{equation}
\tilde{\mathbf{F}} = \mathrm{Attention}(\mathbf{Q}, \mathbf{K}) \times \mathbf{V}.
\vspace{-3pt}
\end{equation}
To facilitate residual fusion and downstream processing, the fused token sequence $\tilde{\mathbf{F}}$ is reshaped and partitioned according to the original token indices of each modality. For the spatially structured modalities (i.e., camera, LiDAR, and radar), the corresponding fused tokens are reshaped back into feature maps and reintegrated into their branches via residual connections. For the GPS modality, which is represented as embedded non-spatial tokens rather than a 2D feature map, the corresponding fused tokens are retained in token form and forwarded to the subsequent fusion stage. In this way, each branch can benefit from cross-modal information while preserving its modality-specific structure.

\vspace{-10pt}
\subsection{Beam Generator and Model Training}
After extracting and fusing multi-scale features, the final representation is aggregated via a softmax-weighted fusion mechanism, as denoted by ${\color{red} \bm{\otimes}}$ in Fig.~\ref{Fig.BeamTransFuser}. Specifically, a learnable importance vector is introduced to adaptively weigh each modality's contribution to the downstream task. This mechanism enables the model to emphasize informative modalities while suppressing noisy or less relevant ones, thereby improving the robustness and discriminative capacity of the fused representation.

Finally, the fused representation $\hat{\mathbf{F}}$ obtained from the softmax-weighted aggregation module is passed through a compact fully connected projection head to produce a low-dimensional embedding suitable for downstream beamforming tasks. Specifically, the projection head consists of a MLP with three linear layers interleaved with ReLU activations. The final output $\hat{\mathcal{Y}}$, i.e., the predicted beam score vector, is given by:
\begin{equation}
\hat{\mathcal{Y}} = \text{MLP}(\hat{\mathbf{F}}) = \mathbf{W}_3 \times \sigma(\mathbf{W}_2 \times \sigma(\mathbf{W}_1 \times \hat{\mathbf{F}})),    
\end{equation}
where $\mathbf{W}_1 \in \mathbb{R}^{512 \times 256}$, $\mathbf{W}_2 \in \mathbb{R}^{256 \times 128}$, and $\mathbf{W}_3 \in \mathbb{R}^{128 \times 64}$ denote the learnable weight matrices corresponding to the three linear layers, respectively. $\sigma(\cdot)$ represents the ReLU activation function.
This dimensionality reduction step serves two purposes: (1) it facilitates computational efficiency by compressing the feature representation and (2) it helps improve task generalization by removing redundant information.

The final output $\hat{\mathcal{Y}} \in \mathbb{R}^{64\times 1}$ represents the class-wise logits, from which the model selects the optimal beam index corresponding to the highest score. To be specific, the chosen beam index is used to retrieve a predefined beamforming vector from a codebook consisting of 64 discrete candidates, which is detailed in Section~\ref{datasets}.
A correct match between the predicted index and the ground-truth index means that the transmitter applies a beamforming vector that aligns well with the dominant propagation path, thereby maximizing signal strength and ensuring efficient communication.
Conversely, an incorrect match leads to beam misalignment, which results in significant signal degradation, increased path loss, and potentially failed communication due to insufficient SNR.

\textbf{Remark:} \textit{In this work, we adopt a 64-dimensional codebook in accordance with the setting of the real-world dataset used for data collection. In principle, the proposed model can be readily extended to massive multiple-input multiple-output (MIMO) scenarios with larger codebooks (e.g., 256-dimensional). However, increasing the codebook size not only enlarges the output dimension, but also may make beam prediction more challenging, since the model needs to distinguish among more closely spaced beam candidates. This may increase beam ambiguity, sensitivity to misalignment, and the overall difficulty of beam prediction, and may also lead to performance degradation if sufficiently discriminative features cannot be learned under a denser codebook. Nevertheless, compared with traditional exhaustive beam search methods, the proposed end-to-end framework still avoids explicit beam-by-beam search over the entire codebook and instead predicts the beam index directly from multimodal observations.}

Building on the above illustration, we are able to construct and train the proposed BeamTransFuser model using a multi-modal dataset collected from real-world scenarios with the gradient descent-based optimizers. The detailed training settings are illustrated in Section~\ref{Simulation Setting}.

\section{Enhancing BeamTransFuser Adaptability via Model Compression and Modality Generation}\label{se:enhancing}
The well-trained BeamTransFuser model can be employed to provide efficient beamforming decisions. However, to effectively learn and fuse multi-modal data, multiple specialized modules are introduced into the BeamTransFuser architecture. While these modules enhance the model's performance, they also significantly increase the number of parameters. This added complexity leads to higher resource consumption and longer inference latency, posing significant challenges for deployment on resource-constrained RSUs.
On the other hand, it is often impractical to equip every RSUs with all types of multi-modal sensors in real-world scenarios. The absence of any expected modality directly causes input dimension mismatch and prevents the model from functioning properly during inference. Consequently, missing modalities introduce substantial challenges for efficient and flexible model deployment in practical environments.

To address the aforementioned challenges, we apply a model compression technique to reduce the size and computational overhead of BeamTransFuser, which is detailed in Section~\ref{compress}. The compressed model not only achieves lower inference latency but also maintains competitive performance.
We then develop a generative model that synthesizes missing modality features based on the available inputs in Section~\ref{Genrative model}. This allows BeamTransFuser to operate reliably under partial observability without requiring re-training.
Together, these enhancements significantly improve the robustness and efficiency of BeamTransFuser, facilitating deployment in dynamic and real-time V2X environments with incomplete sensing and limited edge computing resources.

\vspace{-8pt}
\subsection{Compressing the Learning Model via Splitting Pruning}\label{compress}
\vspace{-2pt}

As aforementioned, the substantial number of model parameters in BeamTransFuser imposes considerable challenges in terms of computing resource consumption and inference latency for the RSU, hindering its practical deployment in real-time V2X scenarios. Therefore, we aim to compress the model into a lightweight version to improve efficiency while retaining performance.
Conventionally, model compression can be achieved through AI techniques such as pruning~\cite{cheng2024survey} and knowledge distillation~\cite{44873}. Pruning removes redundant weights or entire structures (e.g., filters, channels, or layers of neural network) based on certain importance criteria~\cite{cheng2024survey}, while the knowledge distillation transfers the knowledge from a large ``teacher'' model to a smaller ``student'' model by aligning soft targets or intermediate representations~\cite{44873}. However, these two state-of-the-art techniques cannot be directly applied to compress BeamTransFuser due to the following reasons.
\subsubsection{Model Splitting}
\begin{table}[b]
\footnotesize
\centering
\caption{Parameter distribution across model components.}
\label{tab:param_distribution}
\setlength{\tabcolsep}{3pt} 
\begin{tabular}{ccc}
\toprule
\textbf{Module} & \textbf{Param Count (M)} & \textbf{Percentage (\%)} \\
\midrule
Camera Branch   & 21.3   & 27.16 \\
LiDAR Branch    & 11.17  & 14.24 \\
Radar Branch    & 11.18  & 14.25 \\
GPS Branch      & 0.17   & 0.22 \\
Multi-Modal Fusion Blocks & \textbf{34.52} & \textbf{44.03} \\
Other Layers    & 0.08   & 0.10 \\
\midrule
\textbf{Total}  & 78.42  & 100.00 \\
\bottomrule
\end{tabular}
\vspace{-10pt}
\end{table}

On the one hand, the strong interdependence between modality-specific branches and the Transformer-based fusion modules results in tightly coupled representations, making it difficult to identify truly redundant components for pruning without compromising performance. For example, directly applying a global pruning ratio of 90\% to the entire backbone may indiscriminately remove connections across layers with vastly different spatial and semantic characteristics. This often results in disproportionate pruning, where the majority of parameters are removed from large modules (e.g., Layer 4 and Fusion Block 4), while lightweight modules are insufficiently pruned. Such imbalance disrupts the model's structural integrity and leads to noticeable performance degradation. As a result, conventional pruning strategies, which typically rely on a one-size-fits-all approach, are not directly applicable to BeamTransFuser.

On the other hand, applying knowledge distillation to BeamTransFuser would require the design of an additional student architecture for the hierarchical multi-stage fusion backbone. In this setting, the redundancy across different layers, fusion stages, and attention-based components is highly non-homogeneous. Manually designing a student model without prior knowledge of which components are indispensable would introduce additional architecture-design heuristics and may inadvertently suppress critical cross-modal interaction paths~\cite{mansourian2025a}. Therefore, although knowledge distillation is a meaningful compression direction in general, it is not adopted as the main compression route in this work. Instead, we focus on directly compressing the original BeamTransFuser fusion architecture in a data-driven and architecture-aware manner.

Accordingly, we propose a splitting-and-pruning scheme that first decouples the BeamTransFuser architecture into modular components and then applies targeted pruning operations to individual sub-modules. This design enables pruning to be conducted in a more controlled and effective manner, avoiding one-size-fits-all strategies that overlook architectural heterogeneity~\cite{liu2025efficient}. Moreover, it helps minimize interference among tightly coupled branches and preserves the integrity of cross-modal representations. Different from recent pruning work for related multimodal beamforming~\cite{10530437}, which is developed under a distributed federated learning framework and mainly aims to reduce model exchange overhead during distributed training, the proposed scheme is specifically designed to directly compress the model under an inference-oriented deployment objective.
The details of the proposed compression scheme are presented below.

\begin{figure*}[t]
    \centering
    \includegraphics[width=0.88\linewidth]{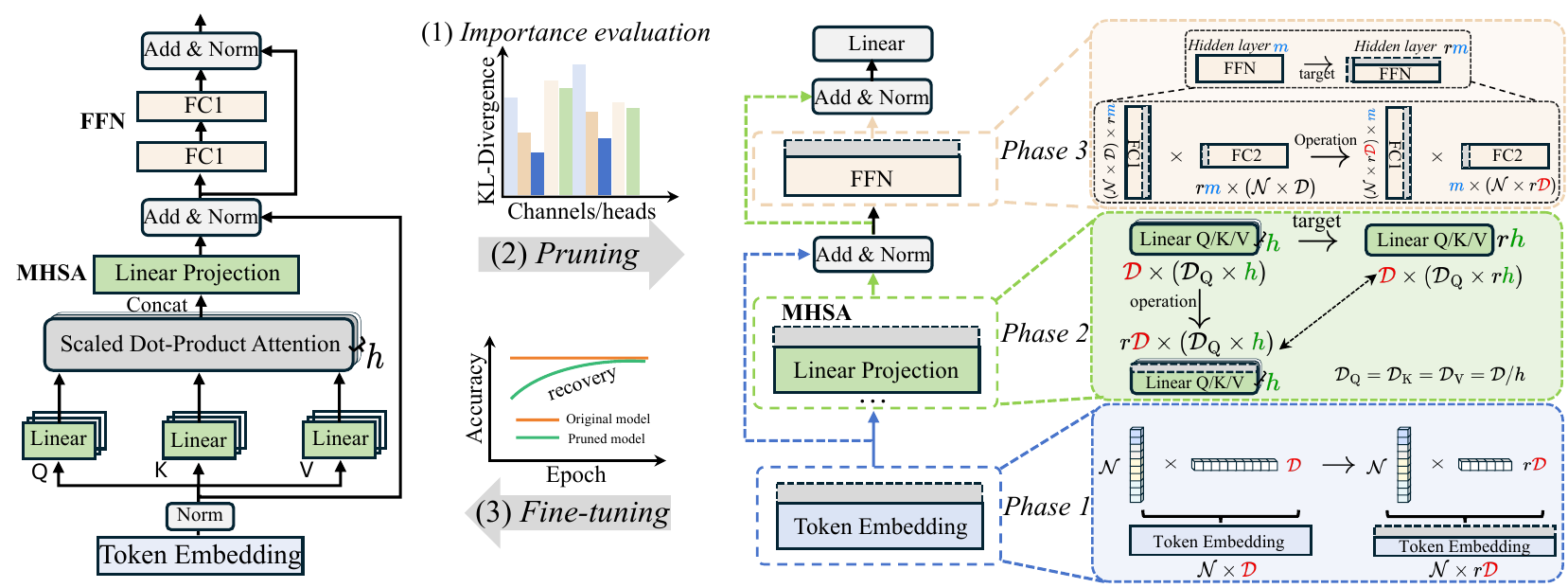}
    \caption{The overview of the multi-modal fusion block and the model compression process. The compression pipeline consists of: (1) KL-divergence-based importance evaluation, (2) progressive pruning, and (3) fine-tuning. In the pruning stage (i.e., the \textbf{right}), Phase~1 reduces the embedding dimension $\mathcal{D}$, while Phase~2 eliminates redundant information without discarding entire attention heads. Phase~3 prunes the hidden neurons in the feed-forward layers and inserts a linear projection layer to restore the original output dimension for compatibility with subsequent blocks.}
    \label{fig:pruning}
\end{figure*} 
As shown in TABLE~\ref{tab:param_distribution}, the majority of parameters in BeamTransFuser are concentrated in the multi-modal fusion blocks (i.e., 44.03\%) and all branches (i.e., 55.82\%). This observation naturally motivates us to compress these modules to reduce the overall model size.
However, although the modality-specific branches (e.g., camera, LiDAR, and radar encoders) contribute a certain proportion of parameters, further compressing these branches would significantly degrade model performance due to their relatively simple and compact ResNet-based structures~\cite{he2016deep}. In contrast, the multi-modal fusion backbone contains multiple Transformer blocks, which offer considerable redundancy and thus provide greater potential for compression without sacrificing task-relevant information.

As a result, we prioritize compressing the Transformer-based fusion blocks while preserving the lightweight structure of the modality-specific encoders. To this end, we decompose the multi-modal fusion backbone of BeamTransFuser into four fusion stages, each containing a multi-modal fusion block. We then apply pruning in a component-wise manner, where the importance of prunable dimensions and channels is evaluated within each stage and the corresponding substructures are compressed accordingly.

\subsubsection{Compressing the Model via Pruning}
We illustrate the pruning process of one splitting component as an example in Fig.~\ref{fig:pruning}. The pruning operation is divided into three stages:  
(1) \textbf{Importance Evaluation}, where the significance of each parameter or structure (e.g., attention heads and hidden neurons) is measured using a predefined criterion;  
(2) \textbf{Progressive Pruning} from Phase 1 to Phase 3, in which low-importance components are gradually removed to avoid drastic performance drops;  
(3) \textbf{Fine-tuning}, where the pruned model is retrained to recover performance.

In the importance evaluation stage, we adopt the Kullback-Leibler (KL) divergence to assess the importance of each structural component by measuring the change in output distribution before and after masking the target structure~\cite{luo2020neural}. Specifically, for each candidate prunable component, we temporarily mask it and compute the KL divergence between the softmax-normalized responses of the self-attention module before and after masking. These responses are taken from the merged attention features prior to the final output projection layer. A higher divergence indicates that removing the corresponding structure causes a more significant shift in the module output behavior, implying greater importance. Conversely, structures with low KL scores are deemed less critical and are pruned in subsequent stages. To reduce sample-specific fluctuations, the KL-based importance score is accumulated over training batches and averaged across the observed data during training.

Subsequently, we are able to prune redundant components within the multi-modal fusion block. To this end, we first analyze the types of parameters that can be pruned. As shown in the right of Fig.~\ref{fig:pruning}, the prunable components include:  
(1) the channels size in residual connections (denoted by the red $\mathcal{D}$ in Phase 1),  
(2) the attention heads in the MHSA modules (denoted by the green $h$), and  
(3) the hidden neurons in the feed-forward network (FFN) layers (denoted by the blue $m$). Specifically, the channel dimension in the residual connections is determined by the feature embedding dimension used in the self-attention module, as defined in (\ref{eq:dimension for embedding}). In other words, this dimension depends on how many embedding channels of  transformer are used to embed the multi-modal tokens~\cite{vaswani2017attention}.
A larger embedding dimension allows the model to capture richer semantic information and represent complex inter-modal interactions more effectively. However, it also introduces significant computational overhead and increases the number of redundant parameters. In contrast, a smaller embedding dimension reduces computational cost and memory consumption, but may limit the model's capacity to represent fine-grained modality-specific or cross-modal features~\cite{cheng2024survey}.

Moreover, attention heads play a fundamental role in the multi-modal fusion block, as
they jointly model cross-modal dependencies within the shared attention
space~\cite{vaswani2017attention}. Nevertheless, they also introduce substantial computational overhead due to the
repeated self-attention operations across heads. As a result, many existing works aim
to reduce the number of attention heads to achieve model compression~\cite{10978953}.
However, in the context of BeamTransFuser, the multi-head self-attention
module operates on a unified token sequence formed by all modalities, rather than relying
on a strictly predefined head-to-modality assignment. Therefore, directly removing entire
attention heads would constitute a coarse-grained structural change that may undesirably
disrupt the learned cross-modal interaction patterns.
For this reason, instead of pruning whole heads, we adopt a more
conservative and finer-grained strategy by reducing the embedding dimensions of the query,
key, and value projections, while preserving the overall multi-head attention structure.
In contrast, the hidden neurons in the FFN layers can be pruned in a more flexible
manner based on the KL scores, as they operate independently on each token and are not
explicitly tied to any particular modality.

Based on the above analysis, it is evident that pruning entire attention heads is non-trivial in our model compression task. Instead of directly removing entire heads from the MHSA module, we adopt a finer-grained strategy by pruning the least important dimensions within the query ($\mathcal{D}_Q$), key ($\mathcal{D}_K$), and value ($\mathcal{D}_V$) projections of the shared attention module across multiple heads. Accordingly, compressing these projections effectively reduces the representation width used inside the MHSA module. This enables us to significantly reduce the number of parameters and computational overhead associated with generating the query, key, and value projections for each head, while maintaining a balanced and expressive attention representation. In other words, we are able to eliminate redundant information without discarding any entire attention head, while preserving the overall multi-head interaction structure. Similarly, the hidden neurons in the FFN can be simplified in a similar fine-grained manner for additional compression.

It is worth noting that pruning reduces the internal representation width inside the MHSA module. To preserve compatibility with the subsequent FFN and later blocks in the pretrained architecture, we reconstruct the corresponding output projection so that the compressed attention representation is mapped back to the original block dimension. The pruned model is then fine-tuned on the original training dataset to recover part of the accuracy loss caused by pruning while retaining the reduced complexity.

We illustrate the entire pruning process with the following example. First, we compute the KL divergence for all candidate pruning components to assess their importance. This results in a list of importance scores for all components. Given a target compression ratio of 10\%, we set the global pruning ratio as $r = 0.1$. Subsequently, pruning is performed within each block according to the corresponding importance scores, while the overall pruning process is coordinated to satisfy the global 10\% compression target. After pruning, we fine-tune the model using the same training dataset to recover potential performance degradation and ensure stable convergence.

\subsection{Filling the Missing Data through Generative Model}\label{Genrative model}
We now introduce our generative model designed to reconstruct missing modality input.
Conceptually, a generative model establishes a mapping from a latent variable space to the data space, enabling the generation of realistic and coherent samples that preserve the semantic and structural characteristics of the training data~\cite{9555209}.

In this work, we develop our generative model based on the variational auto-encoder (VAE) framework~\cite{kingma2013auto}. Compared to other popular generative models such as generative adversarial networks (GANs)~\cite{goodfellow2014generative} and diffusion models~\cite{croitoru2023diffusion}, VAEs offer several advantages.
First, VAEs provide greater training stability and efficiency than GANs, particularly when modeling complex and diverse data distributions~\cite{kingma2019introduction}. Moreover, although diffusion models have demonstrated superior generation quality, they typically require hundreds of iterative denoising steps during inference, leading to latency on the order of seconds~\cite{croitoru2023diffusion}.
Such high latency is incompatible with the real-time requirements of our beamforming task. In contrast, VAEs support generation with significantly lower inference latency~\cite{kingma2013auto}, making them well-suited for delay-sensitive applications such as ours.

Nevertheless, the standard VAE~\cite{kingma2013auto} is inadequate for our task, as it performs unconditional generation without leveraging contextual information from the observed modalities. In other words, the data generated by VAEs is sampled randomly from the overall learned data distribution and may be semantically inconsistent with the actual observation. In contrast, our task requires generating missing modality features conditioned on the available ones, which naturally constitutes a conditional generation problem.

\begin{figure}
    \centering
    \includegraphics[width=0.80\linewidth]{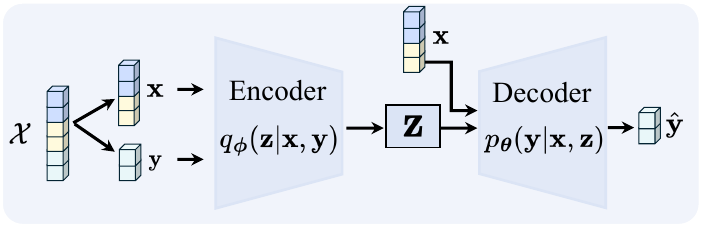}
    \caption{The overview of generative model. The trained decoder can generate data samples from the learned conditional probability distribution \( p_{\boldsymbol{\theta}}(\mathbf{y}|\mathbf{x}) \) using the observed input \( \mathbf{x} \) and sample from $\mathbf{z}$, without requiring the encoder during inference.}
    \label{fig:generative model}
    \vspace{-15pt}
\end{figure}

Consequently, we utilize the conditional variational auto-encoder (CVAE), a variant of VAE that conditions both the encoder and decoder on observed inputs~\cite{NIPS2015_8d55a249}. Let $\mathbf{x}\in \mathcal{X} $ denote the available modalities and $\mathbf{y}\in \mathcal{X}$ represent the missing modality, respectively. In our generative task, we aim to learn a conditional generative model that reconstructs $\mathbf{y}$ based on the $\mathbf{x}$, i.e., to model the conditional probability density function (PDF) $p\left( \mathbf{y}|\mathbf{x} \right) $. 
For example, given the available modalities camera and LiDAR, we aim to leverage them to reconstruct the missing modality (i.e., radar). Under this assumption, the camera and LiDAR data constitute $\mathbf{x}$, and the radar data corresponds to $\mathbf{y}$. Accordingly, the learning objective is to approximate the conditional distribution $p\left( \mathbf{y}|\mathbf{x} \right)$, enabling the model to infer the missing modality from the available modalities. This design allows the model to generate missing data that is semantically consistent with the available modalities. The details are illustrated in Fig.~\ref{fig:generative model} and as follows.

The generative model enhances its representational capacity by introducing a latent variable $\mathbf{z}$, which captures hidden semantic factors that influence the generation process. Accordingly, the conditional PDF is formulated as:
\begin{equation}
p\left( \mathbf{y}|\mathbf{x} \right) = \int_{\mathbf{z}} p\left( \mathbf{y}|\mathbf{x}, \mathbf{z} \right) p\left( \mathbf{z}|\mathbf{x} \right) d\mathbf{z},
\label{eq:pdf-z}
\end{equation}
where $p(\mathbf{z}|\mathbf{x})$ denotes the conditional prior distribution over the latent variable $\mathbf{z}$, and $p(\mathbf{y}|\mathbf{x}, \mathbf{z})$ is the conditional likelihood, representing the probability of generating $\mathbf{y}$ given both the observed input $\mathbf{x}$ and the latent variable $\mathbf{z}$.
In practice, this PDF is typically parameterized by a model $\boldsymbol{\theta }$, such as a deep neural network. Therefore, (\ref{eq:pdf-z}) can be rewritten as:
\begin{equation}
p_{\boldsymbol{\theta }}\left( \mathbf{y}|\mathbf{x} \right) =\int_{\mathbf{z}}{p_{\boldsymbol{\theta }}\left( \mathbf{y}|\mathbf{x},\mathbf{z} \right) p\left( \mathbf{z}|\mathbf{x} \right)d\mathbf{z}.}
\label{eq:pdf}
\end{equation}

The training objective of the generative model is to maximize the conditional log-likelihood $\log p_{\boldsymbol{\theta}}(\mathbf{y}|\mathbf{x})$, which quantifies the likelihood of generating the target modality $\mathbf{y}$ conditioned on the observed inputs $\mathbf{x}$~\cite{kingma2013auto,NIPS2015_8d55a249}. 
Given the optimal generative model parameters ${\boldsymbol{\theta }}^*$, the training process can be formulated as the following optimization problem:
\begin{equation}
\begin{aligned}
 {\boldsymbol{\theta }}^* &= \arg\max_{\boldsymbol{\theta }}\, \log p_{\boldsymbol{\theta }}(\mathbf{y}|\mathbf{x})\\
 &=\mathrm{arg}\max_{\boldsymbol{\theta }} \int_{\mathbf{z}}{p_{\boldsymbol{\theta }}(\mathbf{y}|\mathbf{x},\mathbf{z})\,p\left( \mathbf{z}|\mathbf{x} \right)d\mathbf{z}}.
\end{aligned}
\label{optimation2}
\end{equation}
However, the integral in (\ref{optimation2}) is generally intractable due to the nonlinearity and high dimensionality of the latent variable \( \mathbf{z} \), which makes it difficult to evaluate the conditional marginal likelihood \( p_{\boldsymbol{\theta}}(\mathbf{y}|\mathbf{x}) \) in closed form. To address this challenge, CVAEs introduce an auxiliary neural network (i.e., the encoder) to approximate the true posterior \( p(\mathbf{z} | \mathbf{x}, \mathbf{y}) \) with a tractable variational distribution \( q_{\boldsymbol{\phi}}(\mathbf{z}|\mathbf{x}, \mathbf{y}) \), where \( \boldsymbol{\phi} \) denotes the parameters of the encoder network.

By multiplying and dividing the integrand by the variational distribution \( q_{\boldsymbol{\phi}}(\mathbf{z}|\mathbf{x}, \mathbf{y}) \), the conditional log-likelihood can be expressed as an expectation over this distribution:
\begin{equation}
\begin{aligned}
\log p_{\boldsymbol{\theta}}(\mathbf{y}|\mathbf{x}) 
&= \log \int
\frac{q_{\boldsymbol{\phi}}(\mathbf{z}|\mathbf{x}, \mathbf{y})}{q_{\boldsymbol{\phi}}(\mathbf{z}|\mathbf{x}, \mathbf{y})} 
p_{\boldsymbol{\theta}}(\mathbf{y}|\mathbf{x}, \mathbf{z})\, p(\mathbf{z}|\mathbf{x})\, d\mathbf{z} \\
&= \log \int q_{\boldsymbol{\phi}}(\mathbf{z}|\mathbf{x}, \mathbf{y}) 
\frac{p_{\boldsymbol{\theta}}(\mathbf{y}|\mathbf{x}, \mathbf{z})\, p(\mathbf{z}|\mathbf{x})}{q_{\boldsymbol{\phi}}(\mathbf{z}|\mathbf{x}, \mathbf{y})} \, d\mathbf{z} \\
&= \log \mathbb{E}_{\mathbf{z} \sim q_{\boldsymbol{\phi}}(\mathbf{z}|\mathbf{x}, \mathbf{y})}
\left[
\frac{p_{\boldsymbol{\theta}}(\mathbf{y}|\mathbf{x}, \mathbf{z})\, p(\mathbf{z}|\mathbf{x})}{q_{\boldsymbol{\phi}}(\mathbf{z}|\mathbf{x}, \mathbf{y})}
\right].
\end{aligned}
\label{eq:cvae-loglikelihood}
\end{equation}
Applying Jensen's inequality to the logarithm of the expectation, we have a variational lower bound (ELBO) on the conditional log-likelihood~\cite{kingma2013auto}:
\begin{equation}
\begin{aligned}
&\log p_{\boldsymbol{\theta}}(\mathbf{y}|\mathbf{x}) 
\ge \\
&\mathbb{E}_{\mathbf{z} \sim q_{\boldsymbol{\phi}}(\mathbf{z}|\mathbf{x}, \mathbf{y})}
\left[
\log p_{\boldsymbol{\theta}}(\mathbf{y}|\mathbf{x}, \mathbf{z})
+ \log p(\mathbf{z}|\mathbf{x}) 
- \log q_{\boldsymbol{\phi}}(\mathbf{z}|\mathbf{x}, \mathbf{y})
\right] \\
&= \underbrace{\mathbb{E}_{q_{\boldsymbol{\phi}}} \left[ \log p_{\boldsymbol{\theta}}(\mathbf{y}|\mathbf{x}, \mathbf{z}) \right]}_{\text{Reconstruction Term}} 
\;\; - \;\;
\underbrace{\mathrm{KL} \left( q_{\boldsymbol{\phi}}(\mathbf{z}|\mathbf{x}, \mathbf{y}) \,\|\, p(\mathbf{z}|\mathbf{x}) \right)}_{\text{KL Divergence} \ge 0}.
\end{aligned}
\label{eq:cvae-elbo}
\end{equation}
This expression can be decomposed into two interpretable components: a reconstruction term that encourages accurate generation of the target modality \( \mathbf{y} \), and a KL divergence term that regularizes the approximate posterior to remain close to the prior. In other words, we approximate the conditional log-likelihood \( \log p(\mathbf{y}|\mathbf{x}) \) by maximizing the reconstruction term and minimizing the KL divergence. Accordingly, the following objective is adopted as the base training loss of the CVAE module:
\begin{equation}
\begin{aligned}
\mathcal{L}_{\text{CVAE}}(\boldsymbol{\theta}, \boldsymbol{\phi}) 
&= 
- \mathbb{E}_{\mathbf{z} \sim q_{\boldsymbol{\phi}}(\mathbf{z}|\mathbf{x}, \mathbf{y})}
\left[ \log p_{\boldsymbol{\theta}}(\mathbf{y}|\mathbf{x}, \mathbf{z}) \right] \\
&\quad + \mathrm{KL} \left(
q_{\boldsymbol{\phi}}(\mathbf{z}|\mathbf{x}, \mathbf{y}) \,\|\, p(\mathbf{z}|\mathbf{x})
\right).
\end{aligned}
\label{eq:cvae-loss}
\end{equation}
It is worth noting that explicitly modeling an input-dependent conditional prior \( p(\mathbf{z}|\mathbf{x}) \) typically requires an auxiliary prior network to infer the latent distribution parameters, which introduces additional optimization complexity and increases training and inference overhead. To maintain a lightweight generative module suitable for delay-sensitive V2X edge deployment, following prior works~\cite{kingma2014semi,NIPS2015_8d55a249}, we relax the conditional prior to a fixed isotropic Gaussian distribution, i.e., \( p(\mathbf{z}|\mathbf{x}) = p(\mathbf{z}) = \mathcal{N}(\mathbf{0}, \mathbf{I}) \). Specifically, this simplification only applies to the latent prior and provides a simpler and more stable regularization target for latent-space learning~\cite{kingma2014semi}. The overall framework remains conditional through the variational posterior \( q_{\phi}(\mathbf{z}|\mathbf{x},\mathbf{y}) \) and the generative decoder \( p_{\theta}(\mathbf{y}|\mathbf{x},\mathbf{z}) \). Compared with a learned input-dependent prior \( p(\mathbf{z}|\mathbf{x}) \), this design may offer less flexibility in latent conditional modeling. However, it reduces model complexity and inference overhead, and is therefore adopted as a practical trade-off for deployment-oriented modality completion in the considered scenario. Therefore, the loss function in~(\ref{eq:cvae-loss}) can be efficiently optimized using stochastic gradient descent in conjunction with the reparameterization trick, which enables backpropagation through the sampling process~\cite{NIPS2015_8d55a249}. Moreover, after the base CVAE training stage, we further refine the generative module with a lightweight task-aware fine-tuning step, in which the reconstruction objective is combined with a downstream beam prediction loss. This additional refinement is introduced to enhance the task relevance of the generated modality while preserving its consistency with the target feature space.

As illustrated in Fig.~\ref{fig:generative model}, a trained generative model is capable of reconstructing the missing modality $\hat{\mathbf{y}}$ such that $\hat{\mathbf{y}} \approx \mathbf{y}$. This reconstruction is performed by a decoder parameterized by $\boldsymbol{\theta}$, which takes as input the observed modality $\mathbf{x}$ and a latent noise vector $\mathbf{z}$ sampled from a prior distribution $p(\mathbf{z})$, i.e., $\mathbf{z} \sim p(\mathbf{z})$. 
At inference time, the generation of the missing modality does not require the encoder, as the decoder alone can reconstruct modality features conditioned on the observed input $\mathbf{x}$ and the sampled latent variable $\mathbf{z}$. In this sense, the trained decoder can be regarded as a lightweight generator for missing-modality completion during inference. Moreover, the generated modality features are injected into BeamTransFuser at the feature level, i.e., between the feature extraction module and the multimodal fusion backbone in Fig.~\ref{Fig.BeamTransFuser}. The generated features are then processed jointly with the observed modality features by the hierarchical fusion backbone for subsequent beam inference. In addition, the generative model is trained offline and separately from BeamTransFuser.

\vspace{-5pt}
\section{Performance Evaluation}\label{simultion}
\vspace{-3pt}
\subsection{Dataset}\label{datasets}
\subsubsection{Dataset}
We utilize the DeepSense 6G dataset~\cite{DeepSense2} to evaluate the beam prediction performance of the proposed BeamTransFuser model. As a publicly available and widely used multi-modal dataset, DeepSense 6G provides synchronized multi-modal data collected from real-world vehicular communication scenarios~\cite{DeepSense2}.
To be specific, we focus on the urban street scenario (i.e., the scenarios 31-34 in DeepSense 6G~\cite{DeepSense2}), where a vehicle is equipped with omnidirectional mmWave transmitters and GPS-RTK, while an RSU is equipped with various sensors, including mmWave receivers, an RGB camera, radar, and LiDAR. The dataset comprises the following types of information:
\begin{itemize}
    \item GPS coordinates of the user vehicle,
    \item RGB images, radar, and LiDAR data collected at the RSU, 

    \item A 64×1 power vector recorded at each time step, which represents the received power corresponding to a predefined 64-beam codebook and serves as the basis for selecting the optimal beam index, i.e., the labels $\mathcal{Y}$ defined in (\ref{eq:beamtransfuser}).
\end{itemize}
The data were collected under realistic traffic conditions on College Avenue, a two-way urban street approximately 13 meters wide with bidirectional vehicle flow~\cite{DeepSense2}. After integrating the development and adaptation datasets, we obtain a total of 11,243 data samples (i.e., $N$=11,243). We then split the dataset into 10,118 (i.e., 90\%) samples for training and 1,125 (i.e., 10\%) samples for validation, ensuring that the validation set includes samples from different scenarios to evaluate the model's generalization performance.

Our choice of this dataset is motivated by the following reasons. First, this scenario is particularly representative of real-world V2X environments, as it includes both line-of-sight (LoS) and NLoS conditions, frequent occlusions caused by surrounding vehicles, and complex multipath effects resulting from buildings and traffic dynamics~\cite{DeepSense2}. Moreover, the dataset encompasses multiple scenarios under varying lighting and environmental conditions. Specifically, Scenarios 31 and 32 are collected during the daytime, while Scenarios 33 and 34 are recorded at night. 
These factors pose significant challenges for AI-assisted multi-modal beamforming task, making the dataset well-suited for evaluating the robustness and adaptability of our proposed multi-modal architecture.

\subsection{Experimental Setting}\label{Simulation Setting}
\subsubsection{BeamTransFuser Settings}The detailed architecture of BeamTransFuser is illustrated in Section~\ref{BeamTransFuser Framework}. Specifically, the convolutional layer in each encoder adopts a kernel size of $7 \times 7$ with a stride of 2.
The multi-modal fusion blocks are configured with 4 attention heads, an FFN expansion ratio of 4, and 8 layers. Moreover, the model is trained for 30 epochs with a learning rate of $1\times10^{-4}$. The loss function is focal loss~\cite{lin2017focal}. Furthermore, the embedding dimensions (i.e., $\mathcal{D}$) for each multi-modality fusion block are 64, 128, 256, and 512, respectively.
\subsubsection{Pruning and Generative Model Settings}
We fine-tune the model for ten epochs using the same training data and loss function after each pruning step. Moreover, the encoder of generative model consists of a 2D convolutional layer followed by a ReLU activation, adaptive average pooling, flattening, and two fully connected layers to compute the mean and log-variance of the latent distribution. The dimension of the latent space (i.e., $\mathbf{z}$) is set to 128.

\subsubsection{Evaluation Metrics}
Similar to other related works~\cite{a2022_deepsense,10683225,tian2023multimodal,10.1145/3653644.3680497,10577431,10912462,10949645,10735366} and the evaluation protocol of the DeepSense 6G dataset~\cite{DeepSense2}, we adopt the Distance-based Accuracy Score (DBA-Score) and Top-$k$ accuracy to evaluate the model performance~\cite{DeepSense2}. Specifically, the DBA-Score measures the spatial distance between the predicted and ground-truth beam indices, providing a fine-grained assessment of beam prediction accuracy. The Top-$k$ accuracy calculates the percentage of samples for which the correct beam index is among the top-$k$ predicted candidates~\cite{a2022_deepsense}. Notably, higher DBA-Score and Top-$k$ accuracy indicate more accurate beam alignment, which enhances the received signal quality and ultimately improves the overall communication performance.
\subsubsection{Baselines}We include all existing methods that utilize the same dataset (i.e., scenarios 31 to 34 from the DeepSense 6G) to be the baselines, ensuring a fair and comprehensive comparison. These baselines include: Avatar~\cite{a2022_deepsense}, the position-based and multi-modal models from~\cite{10683225}, TII~\cite{tian2023multimodal}, CMDF~\cite{10.1145/3653644.3680497}, QTNs~\cite{10577431}, ICMFE~\cite{10912462}, MMDL~\cite{10949645}, and MMT~\cite{10735366}.

\subsection{Experimental Results}
\subsubsection{Beamforming Performance via DBA-score}We illustrate the performance of BeamTransFuser in this subsection. To this end, we first present the DBA-score comparison in TABLE~\ref{tab:DBA-modal}, which evaluates multiple multi-modal-based schemes across four urban scenarios (i.e., scenarios 31 to 34).

\begin{table}[b]
\centering
\caption{DBA-score comparison with other multi-modal-based schemes}
\label{tab:DBA-modal}
\begin{threeparttable}
\renewcommand{\arraystretch}{0.9} 
\resizebox{\linewidth}{!}{
\begin{tabular}{cccccc}
    \toprule
    \textbf{Scheme} & \textbf{Overall} & \textbf{S31} & \textbf{S32} & \textbf{S33} & \textbf{S34} \\
    \midrule
    Avatar~\cite{a2022_deepsense} & 0.7162 & 0.6536 & 0.7074 & 0.8576 & 0.7120 \\
    \cite{10683225} & -- & -- & 0.8906 & -- & -- \\
    TII~\cite{tian2023multimodal} & 0.7844 & 0.7298 & 0.7852 & 0.8462 & 0.8433 \\
    CMDF~\cite{10.1145/3653644.3680497} & 0.8910 & -- & -- & -- & -- \\
    QTNs~\cite{10577431} & -- & 0.7605 & 0.8707 & 0.8864 & 0.9124 \\
    ICMFE~\cite{10912462} & 0.8969 & 1.0000 & 0.9020 & 0.8874 & 0.9074 \\
    \textbf{BeamTransfuser (Ours)} & \textbf{0.9129} & \textbf{1.0000} & \textbf{0.9038} & \textbf{0.8988} & \textbf{0.8945} \\
    \bottomrule
\end{tabular}
}
\begin{tablenotes}
    \footnotesize
    \item -- denotes data not reported in the corresponding reference.
\end{tablenotes}
\end{threeparttable}
\end{table}

\begin{table*}[t]
\centering
\caption{Top-$k$ prediction accuracy for each scenario across different schemes}
\label{tab:topk_grouped}
\resizebox{\linewidth}{!}{
\begin{tabular}{c|ccc|ccc|ccc|ccc|c}
    \toprule
    \textbf{Scheme} 
    & \multicolumn{3}{c|}{\textbf{Scenario 31}} 
    & \multicolumn{3}{c|}{\textbf{Scenario 32}} 
    & \multicolumn{3}{c|}{\textbf{Scenario 33}} 
    & \multicolumn{3}{c|}{\textbf{Scenario 34}} 
    & \textbf{Overall} \\
    
    & Top-1 & Top-2 & Top-3 
    & Top-1 & Top-2 & Top-3 
    & Top-1 & Top-2 & Top-3 
    & Top-1 & Top-2 & Top-3 
    & Avg. \\
    
    \midrule

    Position-based~\cite{10683225} 
    & -- & -- & -- 
    & 46.05\% & -- & 79.07\% 
    & -- & -- & -- 
    & -- & -- & -- 
    & 62.51\% \\

    Multi-modal~\cite{10683225} 
    & -- & -- & -- 
    & 53.85\% & -- & 88.07\% 
    & -- & -- & -- 
    & -- & -- & -- 
    & 70.96\% \\


    QTNs~\cite{10577431} 
    & -- & -- & -- 
    & -- &-- & -- 
    & -- & -- & -- 
    & -- & -- & -- 
    & 61.49\% \\

    ICMFE~\cite{10912462} 
    & -- & -- & -- 
    & -- & -- & -- 
    & -- & -- & -- 
    & -- & -- & -- 
    & 76.60\% \\

    MMDL~\cite{10949645} 
    & 33.58\% & -- & -- 
    & 42.82\% & -- & -- 
    & 32.21\% & -- & -- 
    & 41.97\% & -- & -- 
    & -- \\
    
    MMT~\cite{10735366} 
    & -- & -- & -- 
    & 59.5\% & -- & 81\% 
    & -- & -- & -- 
    & -- & -- & -- 
    & 70.25\% \\

    \textbf{BeamTransfuser (Ours)} 
    & \textbf{99.57\%} & \textbf{100\%} & \textbf{100\%} 
    & \textbf{60.72\%} & \textbf{78.98\%} & \textbf{86.20\%} 
    & \textbf{58.78\%} & \textbf{77.82\%} & \textbf{87.65\%} 
    & \textbf{49.14\%} & \textbf{73.72\%} & \textbf{87.59\%} 
    & \textbf{77.27\%} \\
    
    \bottomrule
\end{tabular}
}
\begin{tablenotes}
    \footnotesize
    \item -- denotes data not reported in the corresponding reference.
\end{tablenotes}
\end{table*}
As shown in TABLE~\ref{tab:DBA-modal}, BeamTransFuser achieves the highest overall DBA-score of 0.9129, outperforming all baselines. It also delivers consistently strong performance across all four scenarios, with individual scores of 1.0000 (scenario 31), 0.9038 (scenario 32), 0.8988 (scenario 33), and 0.8945 (scenario 34). This consistency demonstrates the model's robustness under diverse conditions, including variations in lighting (daytime vs. nighttime) and propagation environments (LoS vs. NLoS) where occlusions and multipath effects are common.
Interestingly, several baselines (e.g., Avatar and TII) exhibit noticeable performance gains from Scenario 32 (daytime) to Scenario 33 (nighttime), with improvements of 0.15 and 0.061, respectively. These trends indicate that certain nighttime conditions, such as reduced visual clutter or more uniform illumination, may benefit some models. However, the degree of improvement varies significantly, likely due to differences in modality dependencies and fusion strategies.

In contrast, BeamTransFuser maintains stable performance across Scenarios 32 to 34, indicating strong invariance to environmental changes. 
It is also worth noting that although BeamTransFuser is slightly outperformed by QTNs (0.9124) and the method in~\cite{10912462} (0.9074) in Scenario 34, with performance differences of only 1.99\% and 1.42\%, respectively, BeamTransFuser achieves the highest overall DBA-score of 0.9129 among all methods. This demonstrates that while other models may excel in isolated scenarios, BeamTransFuser offers more consistent and superior performance across diverse environments, highlighting its strong generalization and robustness.

\subsubsection{Beamforming Performance via Top-$k$}
We then show the Top-$k$ accuracy in TABLE~\ref{tab:topk_grouped}. Specifically, as several baseline methods do not provide complete scenario-wise or Top-$k$ results, this poses challenges for direct comparison. To address this, we apply the following strategy. For methods with partial reports, such as the Position-based and Multi-modal schemes in~\cite{10683225}, and MMT~\cite{10735366}, we compute their overall accuracy by averaging the available Top-$k$ values. For instance, in Scenario 32, where only Top-1 and Top-3 accuracies are reported, we take their mean to approximate the overall score. Since Top-2 typically lies between Top-1 and Top-3, excluding it is unlikely to introduce significant bias, especially under the assumption of monotonic accuracy trends.

Furthermore, the MMDL method~\cite{10949645} only reports Top-1 accuracy for each scenario, without providing Top-2 or Top-3 values. To avoid misleading interpretation due to this limited granularity, we do not report its overall accuracy. Regarding ICMFE~\cite{10912462}, although its original paper does not explicitly present the performance for Scenario 31, it claims perfect accuracy (i.e., 100\%) in that case. Along with the reported accuracies of 69.95\%, 66.35\%, and 70.11\% for Scenarios 32, 33, and 34, respectively, we compute an overall accuracy of 76.60\% for fair inclusion in the comparison~\cite{10912462}. This operation enables a more consistent and transparent comparison across all evaluated schemes.

As shown in Table~\ref{tab:topk_grouped}, BeamTransfuser achieves consistently high Top-$k$ prediction accuracy across all four scenarios, outperforming all baseline methods. In Scenario~31, it reaches 99.57\% Top-1 accuracy and 100\% for both Top-2 and Top-3
In Scenario~32, BeamTransfuser achieves 60.72\% Top-1 accuracy, which increases to 78.98\% and 86.20\% for Top-2 and Top-3, respectively. These results consistently surpass other baselines, demonstrating the model's ability to rank correct beam candidates among the top predictions even when the top-1 prediction is suboptimal.
Moreover, for the nighttime scenarios (i.e., Scenarios~33 and 34), BeamTransfuser exhibits similar trends, with Top-1 accuracies of 58.78\% and 49.14\%, and Top-3 accuracies of 87.65\% and 87.59\%, respectively.

Compared to other methods that either lack complete Top-$k$ evaluations or show greater performance variability (e.g., MMDL shows lower Top-1 scores across scenarios, while QTNs does not provide scenario-specific details), BeamTransfuser offers not only a higher overall Top-$k$ accuracy (77.27\%) but also more consistent results across scenarios and metrics.

The superior performance of BeamTransFuser is attributed to its transformer-based multi-modal fusion mechanism, which effectively captures cross-modal dependencies while preserving modality-specific information. In addition, its hierarchical multi-modal fusion blocks and early-stage residual connections enhance both low-level structural features and high-level semantic representations, enabling the model to adapt effectively to complex and noisy urban V2X environments.

\subsubsection{Model Performance after Pruning}
\begin{figure}[t]
	\centering
	\begin{subfigure}[b]{0.5\linewidth}
		\centering
		\includegraphics[width=1.0\linewidth]{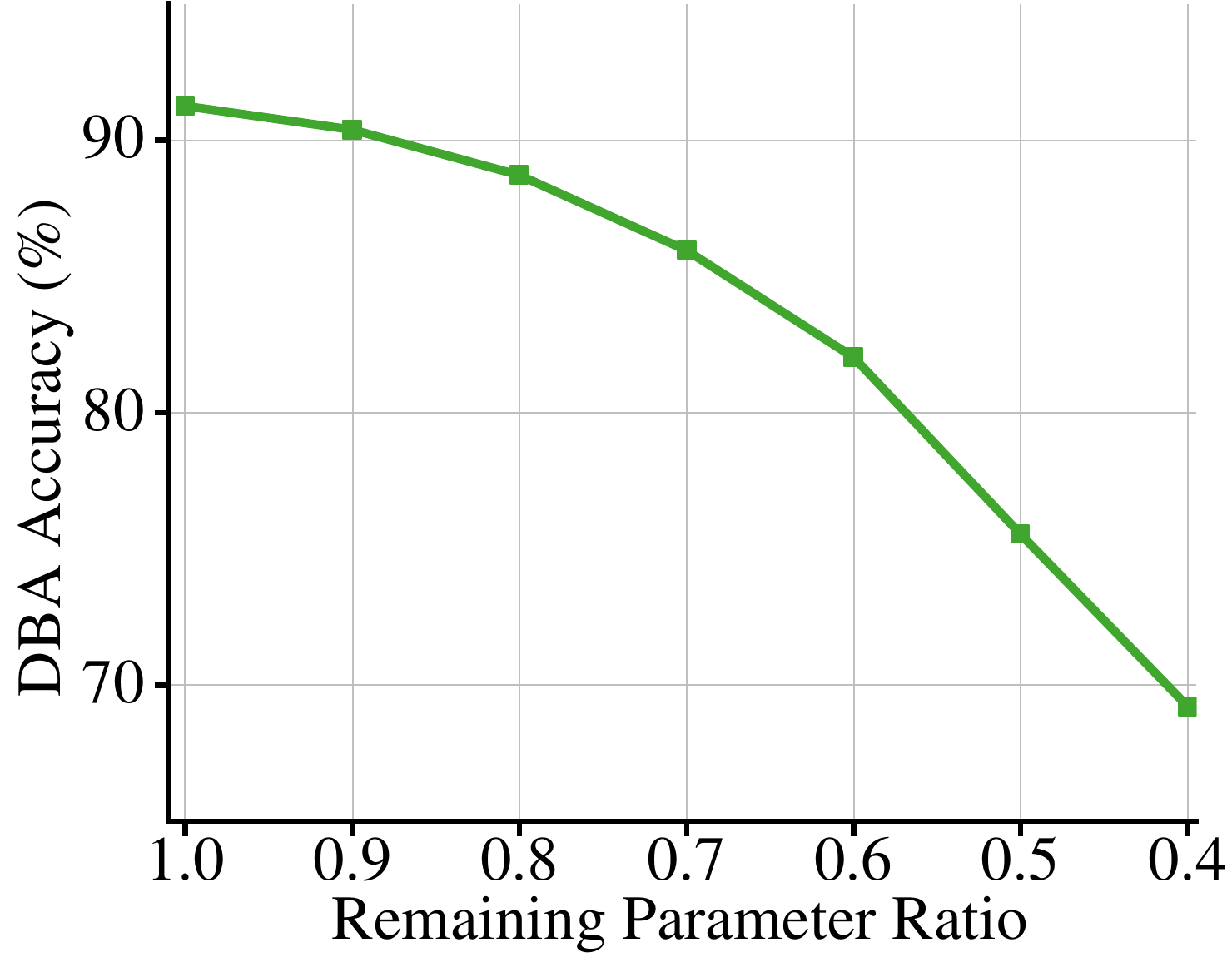}
		\caption{}
	\end{subfigure}%
	\begin{subfigure}[b]{0.5\linewidth}
		\centering
		\includegraphics[width=1.0\linewidth]{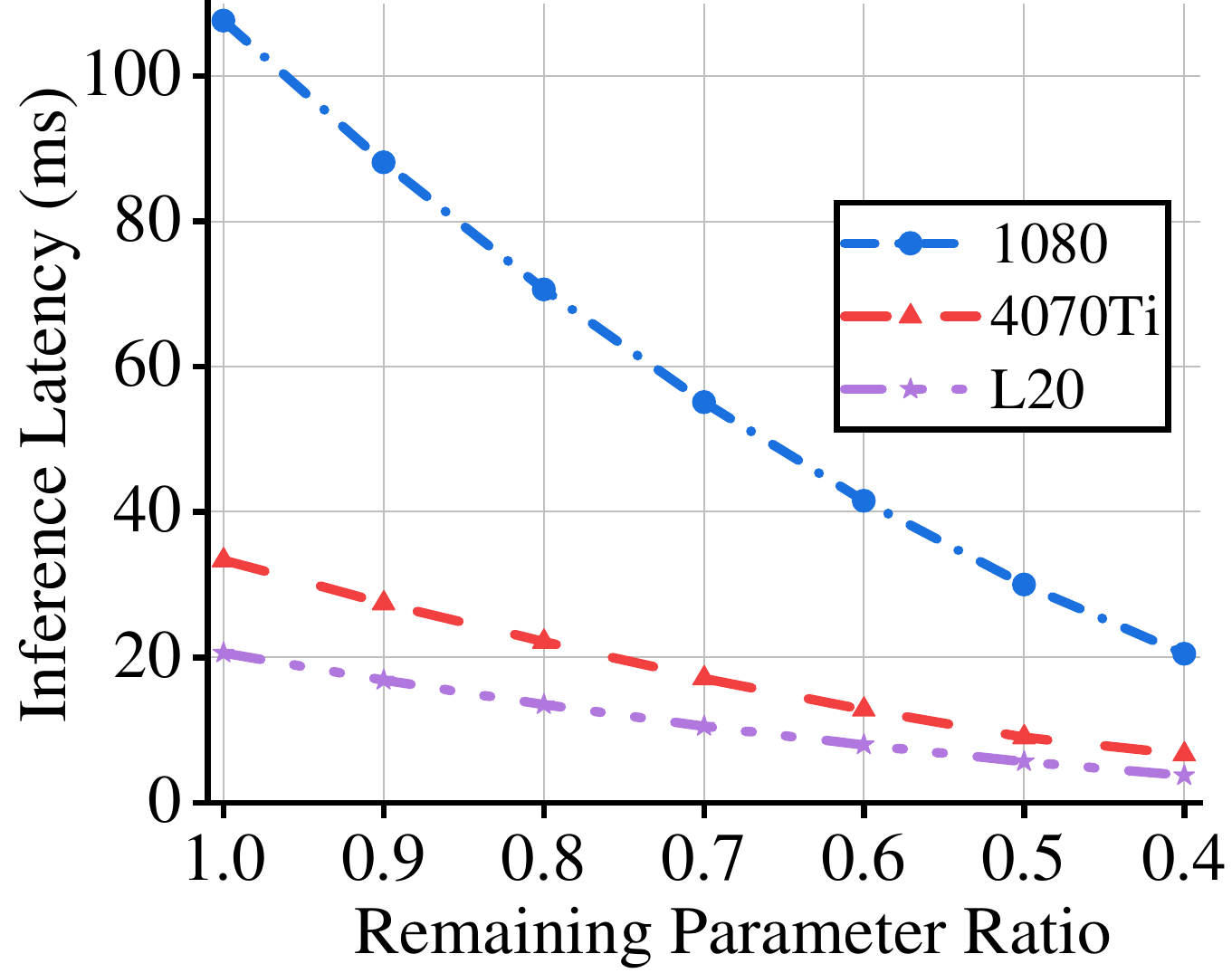}
		\caption{}
	\end{subfigure}
	\caption{(a) DBA accuracy vs. remaining parameter ratio and (b) inference latency vs. remaining parameter ratio on various devices.}
	\label{fig:compression.}
    \vspace{-13pt}
\end{figure}

We show the inference latency and accuracy under varying remaining parameter ration of Transformer modules (i.e., $1-r$) in Fig.~\ref{fig:compression.}.
As shown in Fig.~\ref{fig:compression.}(a), the DBA accuracy gradually declines as the compression ratio increases. Specifically, when the remaining parameter ratio decreases to 0.9 and 0.8, the accuracy slightly drops to 90.38\% and 88.71\%, respectively. In other words, reducing approximately 10\% and 20\% of the model's 78.42 million parameters through our tailored splitting-based pruning strategy has negligible impact on overall performance. 
With increasingly aggressive compression (i.e., lower parameter retention), the accuracy degrades to 85.95\% and 82.02\% at retention ratios of 0.7 and 0.6, and further declines to 75.54\% and 69.21\% at 0.5 and 0.4, respectively.
Notably, even with a remaining parameter ratio of 0.6, the compressed model still achieves competitive performance compared to existing baseline methods such as~\cite{a2022_deepsense,tian2023multimodal,10577431}.

We further evaluate the inference latency across three hardware platforms: GTX 1080 (legacy consumer grade), RTX 4070Ti (mainstream consumer grade), and L20 (data center grade GPU). As shown in Fig.~\ref{fig:compression.}(b), the inference latency consistently decreases with lower remaining parameter ratios, achieving over 80\% reduction at a 0.4 ratio across all platforms. At a ratio of 0.7 (DBA 85.95\%), latency on mainstream consumer-grade devices drops below 20 ms. At 0.6 (DBA 82.02\%), all platforms maintain latency below the 40 ms threshold defined by 5GAA TR P-170142, enabling real-time deployment in 5G-V2X scenarios~\cite{5GAA_P190033}.

\subsubsection{Generative Model Performance}
To evaluate the performance of the generative model, we conduct a modality masking experiment. Specifically, we simulate the absence of a modality by replacing its original input with either zero-filled features or random Gaussian noise (with mean zero and standard deviation one), while keeping the remaining modalities intact. The corrupted inputs are then fed into BeamTransFuser for inference. This experimental setup enables a quantitative assessment of the performance degradation caused by missing input sources under different baseline assumptions. Furthermore, we compare these degraded results with the performance obtained when the missing modality is replaced by features generated from our trained generative model (i.e., the CVAE), thereby highlighting the effectiveness of generative completion in restoring beam prediction performance. In addition, we also compare with training-time missing-modality learning, which serves as a robust retraining baseline by exposing the beam prediction model to incomplete sensing inputs during training~\cite{CHOI2019259}. Specifically, this baseline is implemented by applying modality dropout with a probability of 50\% during training, where either radar or LiDAR is randomly selected and replaced with zero-valued features when dropout is triggered. The trained model is then evaluated under both full-data and missing-modality conditions. It is worth noting that the GPS modality is not generated in our framework, since it is inherently provided by the vehicle's onboard systems.

As illustrated in TABLE~\ref{tab:missing_modality_single}, when the radar modality is missing, the beam prediction accuracy drops substantially under both the zero-filled and Gaussian-noise baselines, with the Top-1 accuracy decreasing to 6.67\% under both settings. In contrast, when the missing radar features are replaced by CVAE-generated features, the Top-1, Top-2, and Top-3 accuracies recover to 45.51\%, 70.31\%, and 82.58\%, respectively. A similar trend is observed for LiDAR. Its absence also reduces the Top-1 accuracy to 6.67\% under both baselines, whereas the generated LiDAR features improve the Top-1, Top-2, and Top-3 accuracies to 56.18\%, 79.82\%, and 89.07\%, respectively. These results demonstrate that the proposed generative model can effectively reconstruct useful missing-modality representations and substantially recover downstream beam prediction performance. Moreover, we also measure the latency overhead introduced by CVAE-based modality completion on the NVIDIA L20 platform. Specifically, the additional overhead of CVAE-based completion is \(17.11 \pm 0.05\) ms for missing radar and \(16.73 \pm 0.34\) ms for missing LiDAR. When combined with the original unpruned BeamTransFuser backbone, the complete missing-modality recovery pipeline takes \(36.27 \pm 0.04\) ms and \(36.32 \pm 0.09\) ms, respectively, remaining within the 40 ms latency requirement reported by 5GAA for latency-sensitive V2X applications~\cite{5GAA_P190033}.

We then compare the proposed method with the training-time modality-dropout baseline. As shown in TABLE~\ref{tab:missing_modality_single}, this baseline improves the missing-modality results compared with zero-filled and Gaussian-noise replacement by retraining the beam prediction model with incomplete inputs. However, this robustness is obtained by sacrificing part of the full-modality prediction accuracy: the full-data Top-1, Top-2, and Top-3 accuracies decrease from 60.94\%, 78.43\%, and 86.27\% to 54.67\%, 70.53\%, and 79.19\%, respectively. In contrast, the proposed CVAE-based completion preserves the original full-modality BeamTransFuser and activates the generative module only when a sensing modality is missing, thereby retaining the full-modality prediction pipeline while recovering missing-modality information.

\begin{table}[t]
\centering
\caption{Top-\(k\) beam prediction accuracy under full-data and missing-modality conditions. ``Miss.'' denotes the missing modality, ``Modality dropout'' refers to the training-time missing-modality baseline, and ``Gen'' denotes CVAE-based reconstruction of the missing modality using the two available modalities.}
\label{tab:missing_modality_single}
\footnotesize
\setlength{\tabcolsep}{3pt}
\renewcommand{\arraystretch}{1.05}
\begin{tabular}{c c c c c}
    \toprule
    \textbf{Scenario} & \textbf{Method / Condition} & \textbf{Top-1 (\%)} & \textbf{Top-2 (\%)} & \textbf{Top-3 (\%)} \\
    \midrule
    \multirow{4}{*}{Miss. Radar}
      & Zero-filled      & 6.67  & 9.07  & 15.02 \\
      & Gaussian noise   & 6.67  & 9.42  & 14.22 \\
      & Modality dropout & \textbf{54.92} & \textbf{71.10} & 79.56 \\
      & Gen (Cam+LiDAR)  & 45.51 & 70.31 & \textbf{82.58} \\
    \midrule
    \multirow{4}{*}{Miss. LiDAR}
      & Zero-filled      & 6.67  & 9.07  & 13.96 \\
      & Gaussian noise   & 6.67  & 9.07  & 13.78 \\
      & Modality dropout & 51.85 & 65.52 & 74.11 \\
      & Gen (Cam+Radar)  & \textbf{56.18} & \textbf{79.82} & \textbf{89.07} \\
    \midrule
    \multirow{2}{*}{Miss. Camera}
      & Zero-filled      & 2.40  & 4.27  & 6.40  \\
      & Gaussian noise   & 3.02  & 6.31  & 8.27  \\
    \midrule
    \multirow{2}{*}{Full Data}
      & BeamTransFuser & \textbf{60.94} & \textbf{78.43} & \textbf{86.27} \\
      & Modality dropout & 54.67 & 70.53 & 79.19 \\
    \bottomrule
\end{tabular}
\end{table}

It is also worth noting that the zero-filled and Gaussian-noise baselines lead to highly similar degradation in our experiments. The reason is that, although Gaussian noise introduces stronger perturbation than zero filling, neither of them provides meaningful structural or semantic information for the missing modality. Therefore, from the perspective of the downstream fusion backbone, both settings effectively correspond to the absence of task-relevant modality information, which explains why they yield similar beam prediction performance. In this sense, zero-filled replacement can be regarded as a more direct missing-modality baseline, whereas Gaussian-noise replacement serves as a corruption-based robustness test.

For the camera modality, the performance degradation is even more severe, with the Top-1 accuracy dropping to 2.40\% and 3.02\% under zero-filled and Gaussian-noise replacement, respectively, highlighting the critical role of visual information in BeamTransFuser. Notably, we do not display the generated accuracy for the missing camera case due to the following limitations.
First, RGB images contain rich and high-dimensional semantic information (i.e., 256$\times$256$\times$3), which poses a significant challenge for a conditioned generative model. The high resolution and fine-grained content of RGB images demand detailed spatial and contextual understanding, making it difficult to accurately reconstruct them from sparse modalities such as LiDAR and radar. Moreover, generating full-resolution image features introduces significant computational overhead, which is undesirable in time-sensitive communication scenarios~\cite{croitoru2023diffusion}. Therefore, we refrain from applying generative compensation to the camera modality in this work. This limitation further motivates our future work on extending the generative model to support camera modality reconstruction with acceptable generative quality and inference latency.

\section{Conclusion}\label{conclusion}
In this work, we have proposed a robust and adaptive multi-modal beamforming system for real-world V2X networks. Specifically, we have designed a unified end-to-end deep learning framework, BeamTransFuser, which extracts complementary spatial and semantic features from heterogeneous sensing modalities and fuses them via hierarchical Transformer modules to enable accurate and context-aware beam prediction. To support efficient deployment on RSUs, we have developed a splitting-based model compression strategy that tailors pruning to the structural characteristics of each module, achieving substantial parameter reduction with negligible performance degradation. Furthermore, to address the challenge of missing sensor modalities in practical scenarios, we have integrated a generative compensation mechanism capable of reconstructing absent modality features from available inputs, thereby enhancing model robustness without requiring retraining. Extensive evaluations on real-world datasets demonstrate that BeamTransFuser outperforms existing methods in terms of accuracy, latency, and adaptability, underscoring its potential to facilitate intelligent and resilient beamforming in future 6G-enabled V2X environments.

\bibliographystyle{IEEEtran}
\bibliography{bibRef}

\end{document}